\documentclass[iop]{emulateapj}

\usepackage{graphicx}
\usepackage{dcolumn}
\usepackage{bm}
\usepackage{epsf}
\usepackage{verbatim}
\usepackage{amsmath}
\usepackage{amsfonts}
\usepackage{pstricks}
\usepackage{natbib}

\def\myfig#1{#1}

\newcommand{\cm}{\mbox{ cm}}
\newcommand{\km}{\mbox{ km}}
\newcommand{\pc}{\mbox{ pc}}
\newcommand{\kpc}{\mbox{ kpc}}
\newcommand{\Mpc}{\mbox{ Mpc}}
\newcommand{\se}{\mbox{ s}}
\newcommand{\yr}{\mbox{ yr}}

\newcommand{\eV}{\mbox{ eV}}
\newcommand{\keV}{\mbox{ keV}}
\newcommand{\muG}{\mbox{ $\mu$G}}
\newcommand{\pr}{\partial}
\newcommand{\grad}{\bm{\nabla}}
\newcommand{\vect}[1]{\mathbf{#1}}
\newcommand{\fin}{\mbox{ .}}
\newcommand{\coma}{\mbox{ ,}}
\newcommand{\ie}{\emph{i.e.} }
\newcommand{\eg}{\emph{e.g.,} }

\newcommand{\myF}{{\mathcal{F}}}
\newcommand{\mydu}{\Delta u_2}

\newcommand{\myr}{{r'}}
\newcommand{\hatb}[1]{\mathbf{\hat #1}}
\newcommand{\unit}[1]{\bm{\hat{#1}}}
\newcommand{\myhatbr}{\hatb{r}'}

\newcommand{\myth}{{\theta}}
\newcommand{\myph}{{\phi}}
\newcommand{\myPth}{{P_{th}}}
\newcommand{\myPNT}{{P_{nt}}}

\newcommand{\mypP}{{\lambda}}

\newcommand{\myq}{{q}}

\newcommand{\myin}{{i}}
\newcommand{\myout}{{o}}
\newcommand{\mach}{{\mathcal{M}}}

\newcommand{\myl}{{l}}

\shorttitle{Dynamics and magnetization in galaxy cluster cores}
\shortauthors{Keshet et al.}

\begin{document}

\title{Dynamics and Magnetization in Galaxy Cluster Cores Traced by X-ray Cold Fronts}

\author{Uri Keshet\altaffilmark{1}, Maxim Markevitch, Yuval Birnboim, and Abraham Loeb}

\affil{Harvard-Smithsonian Center for Astrophysics, 60 Garden St.,
  Cambridge, MA 02138, USA}

\altaffiltext{1}{Einstein fellow}

\begin{abstract}
Cold fronts (CFs) --- density and temperature plasma discontinuities --- are ubiquitous in cool cores of galaxy clusters, where they appear as X-ray brightness edges in the intracluster medium, nearly concentric with the cluster center.
We analyze the thermodynamic profiles deprojected across core CFs found in the literature.
While the pressure appears continuous across these CFs, we find that all of them require significant centripetal acceleration beneath the front.
This is naturally explained by a tangential, nearly sonic bulk flow just below the CF, and a tangential shear flow involving a fair fraction of the plasma beneath the front.  Such shear should generate near-equipartition magnetic fields on scales $\lesssim 50\pc$ from the front, and could magnetize the entire core.
Such fields would explain the apparent stability of cool-core CFs and the recently reported CF--radio minihalo association.
\end{abstract}


\maketitle


\section{Introduction}

In the past decade, high resolution X-ray observations have revealed an abundance of density and temperature discontinuities known as cold fronts (CFs).
They are broadly classified as merger CFs (not discussed here) and cool core CFs;
for review, see \cite{MarkevitchVikhlinin07}.

Core CFs are observed in more than half of the otherwise relaxed, cool core clusters \citep[CCs,][]{MarkevitchEtAl03Proc}, in distances up to $\sim 400\kpc$ from the center.
They are usually nearly concentric or spiral, and multiple CFs are often observed in the same cluster.
The temperature contrast across such a CF is $\myq\equiv T_{\myout}/T_{\myin} \sim 2$ \citep{OwersEtAl09}, where inside/outside subscripts $i/o$ refer to regions closer to/farther from the cluster center, or equivalently below/above the CF.
The plasma beneath the CF is typically denser, colder, lower in entropy and higher in metallicity than the plasma above it.
Usually, little or no pressure discontinuity is identified across these CFs, suggesting negligible CF motion \citep{MarkevitchVikhlinin07}.

Such CFs were modeled as associated with large scale ``sloshing'' oscillations of the intracluster medium (ICM) driven, for example, by mergers \citep[][henceforth M01]{MarkevitchEtAl01}, possibly involving only a dark matter subhalo \citep{TittleyHenriksen05,AscasibarMarkevitch06}, or by weak shocks/acoustic waves displacing cold central plasma \citep{ChurazovEtAl03, FujitaEtAl04}.

We argue that a tangential shear flow across and beneath CFs is ubiquitous in CCs.
One line of argument relies on thermodynamic profiles across CFs found in the literature.
We show that all these profiles reveal a distinct deviation from hydrostatic equilibrium, and argue that this can be naturally interpreted only as shear.
An independent indication for the presence of shear across CC CFs is their remarkable sharpness and stability, which implies strong magnetic fields along the CF plane. In the absence of radial CF motion, the most natural explanation for the persistence of such fields is shear across the CF.

In \S\ref{sec:Deviations} we show that deviations from hydrostatic equilibrium, previously reported in two CC CFs, are common among such CFs, and argue in \S\ref{sec:DifferentialRotationAcrossCF} that this directly gauges bulk tangential flows and shear along and beneath these CFs.
In \S\ref{sec:MagneticAmplification} we show that such shear can magnetize the core, and produce near-equipartition fields along CFs, thus stabilizing them. We summarize and discuss the implications of our model in \S\ref{sec:Discussion}.
We assume a Hubble constant $H=70\km\se^{-1}\Mpc^{-1}$.
Error bars are $1\sigma$ confidence levels.

\section{Deviations from hydrostatic equilibrium}
\label{sec:Deviations}

\begin{figure}[ht]
\centerline{\epsfxsize=10cm \epsfbox{\myfig{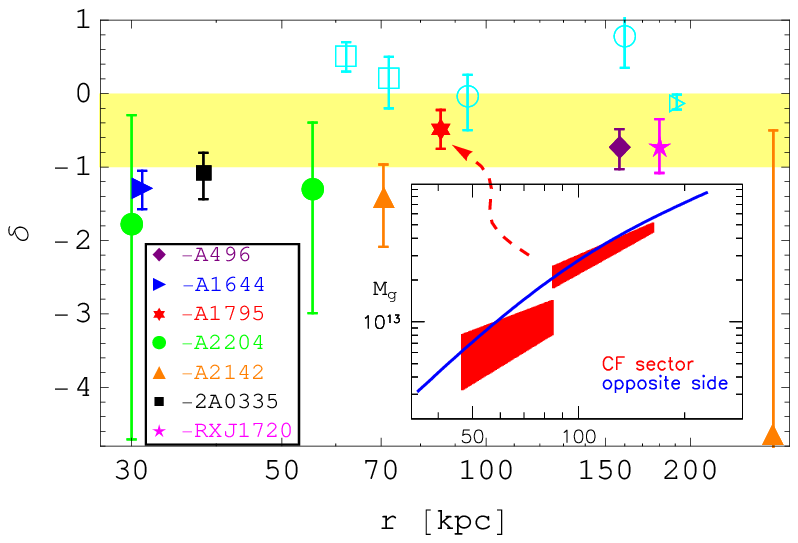}}}
\caption{
Normalized discontinuity $\delta$ in pressure acceleration, plotted against radius $r$ for CFs in various CCs (filled symbols).
Data: \citet[][for A496]{TanakaEtAl06}, \citet[][for A1644]{JohnsonEtAl10}, M01 (for A1795), \citet[][for A2204]{SandersEtAl05}, and \citet[][for 2A0335]{SandersEtAl09}.
Data with $\rho$ but no $T$ deprojection: \citet[][for RX J1720+26]{MazzottaEtAl01} and \citet[][for A2142]{MarkevitchEtAl00}.
The pressure is (consistent with being) continuous across all these CFs.
Also shown are the acceleration discontinuities in some deprojected regions where no CF is present (empty cyan symbols), used as a control sample.
The $-1<\delta<0$ (shaded) region is roughly consistent with subsonic shear and flow beneath the CF, $0<\Delta(u^2) = (u_\myin+u_\myout)\Delta u <c_i^{2}$.
\\
Inset: Gravitating mass (in $M_\odot$) assuming hydrostatic equilibrium, as a function of $r$ (in kpc) in two opposite sectors in A1795, one harboring the CF analyzed (adopted from M01).
\label{fig:CFJump}
}
\end{figure}

The mass density $\rho(r)$ and pressure $P(r)$ radial profiles near CFs in RX J1720+26 and A1795 were reported to be inconsistent with hydrostatic equilibrium \citep[M01;][]{MazzottaEtAl01}. Indeed, assuming such equilibrium, $a_r=-\rho^{-1}\pr_r P - G M_g/r^2=0$, where $a_r$ is the radial acceleration and $G$ is Newton's constant, would imply an unphysical gravitating mass profile $M_g(r)$ that becomes abruptly larger just outside the CF.
In RX J1720, $M_g$ derived from the above assumption jumps by a factor of 5 (at a $2.5\sigma$ confidence level) at the CF, and there is marginal evidence for a pressure discontinuity which translates to radial CF motion with Mach number $\mach=0.4_{-0.4}^{+0.7}$ \citep{MazzottaEtAl01}.
In A1795, $M_g$ jumps by a factor of 2 ($2.5\sigma$) and the pressure appears continuous across the CF (Figure \ref{fig:CFJump}, inset; see M01).

To investigate the prevalence of this phenomenon, we calculate the pressure acceleration discontinuity
\begin{eqnarray}
\Delta a_r \equiv a_{r,i}-a_{r,o} = \Delta (-\frac{\pr_r P}{\rho})=-\frac{k_B}{\mu r}\Delta(\mypP T)
\end{eqnarray}
across various CFs for which deprojected thermodynamic profiles have appeared in the literature. Here we defined $\Delta A\equiv A_{\myin}-A_{\myout}$ (for any quantity $A$) and $\mypP\equiv \pr\ln P/\pr\ln r$ (typically $\mypP<0$), with $k_B$ being the Boltzmann constant, $\mu\simeq 0.6 m_p$ being the average particle mass (assumed constant), and $m_p$ being the proton mass.
The gravitating mass $M_g(r)$ is assumed to be continuous everywhere.

We extract all the thermodynamic profiles across CFs in cool cluster cores from the literature.
We select only profiles that have been deprojected along the line of sight, where at least two radial bins exist on each side of the CF such that $\pr_r P$ can be estimated.
When possible, we omit radial bins which are very far ($>4$ times in radius) from the CF, to avoid excessive contamination.
We end up with six CF profiles in which $\mypP T$ can be calculated based on 2-3 radial bins on each side of the front, supplemented by three CFs for which only the density has been deprojected.
We present the (dimensionless) normalized acceleration discontinuity $\delta\equiv (c_i^2/r)^{-1}\Delta a_r$ in Figure \ref{fig:CFJump}, along with the literature references.
Here, $c=(\Gamma P/\rho)^{1/2}$ is the sound velocity, with $\Gamma=5/3$ being the adiabatic index.

All the $\Delta a_r$ values we find are negative, and inconsistent with zero at confidence levels ranging between $1.1$ and $5.4\sigma$.
The CF data points shown in Figure 1 are consistent with a best fit $\delta=-0.78\pm 0.14$, \ie nonzero at $5.5\sigma$ (the average is not necessarily physically meaningful).
As a sanity check, we apply the exact same procedure to parts of the same deprojected thermodynamic profiles where no CF is present.
This is possible in a few cases, where the profiles extend sufficiently far from the CF, shown in Figure~(\ref{fig:CFJump}) as empty cyan symbols.
These continuum $\delta$ values are consistent with zero or are slightly positive (within $2.5\sigma$ from zero; a resolution effect explained below). Overall, they are best fit by $\delta=+0.07\pm 0.08$.
This indicates that the $a_r$ discontinuity at CFs is real.

Furthermore, the effect is qualitative and robust.
As $T$ jumps by a factor of $\myq\simeq 2$ as one crosses outside the CF, hydrostatic equilibrium would require a slower radial decline of pressure outside the CF, $\mypP_{\myin}/\mypP_{\myout}\simeq \myq$. However, the pressure profile typically \emph{steepens} outside the CF.
Hence, two phenomena contribute to the $\Delta a_r<0$ discontinuity: the density drop, and the pressure steepening.
Neither of these effects can be inverted by spherical deprojection.
The differentiation procedure used to estimate $\mypP$ removes long wavelength errors, so the significance of $\delta$ is probably better than that estimated above. Finally, note that resolution effects introduce an opposite bias, toward positive $\Delta a_r$, as $(-\rho^{-1}\pr_r P)$ typically declines with increasing $r$. This accounts for the slight tendency of continuum regions toward $\Delta a_r>0$.

We conclude that deviations from hydrostatic equilibrium are common in CC CFs, with less pressure support below the CF, $\Delta a_r<0$,
such that if the plasma above the CF is in equilibrium, the plasma below should be falling.
This phenomenon, previously reported in RX J1720 and in A1795, has been interpreted as evidence that the core is sloshing, with plasma near the CF \emph{radially} accelerating in some phase of the oscillation cycle. For the apparently stationary CF in A1795, cool plasma would then be sloshing at its maximal radial displacement, where it has zero radial velocity but nonzero radial acceleration (M01).

We note, however, that in this picture, without additional effects, the discontinuity in radial acceleration $\Delta a_r\equiv a_{r,\myin}-a_{r,\myout} < 0$ would cause an unphysical gap to open along the CF.
A different interpretation of the observations is therefore required.
Below we argue that the most natural interpretation of the data involves a noncontinuous (within the observational resolution) centripetal acceleration of flow along the curved CF.

\section{Tangential shear flow beneath CFs}
\label{sec:DifferentialRotationAcrossCF}

\begin{figure}[h]
\vspace{5mm}
\centerline{\epsfxsize=7cm \epsfbox{\myfig{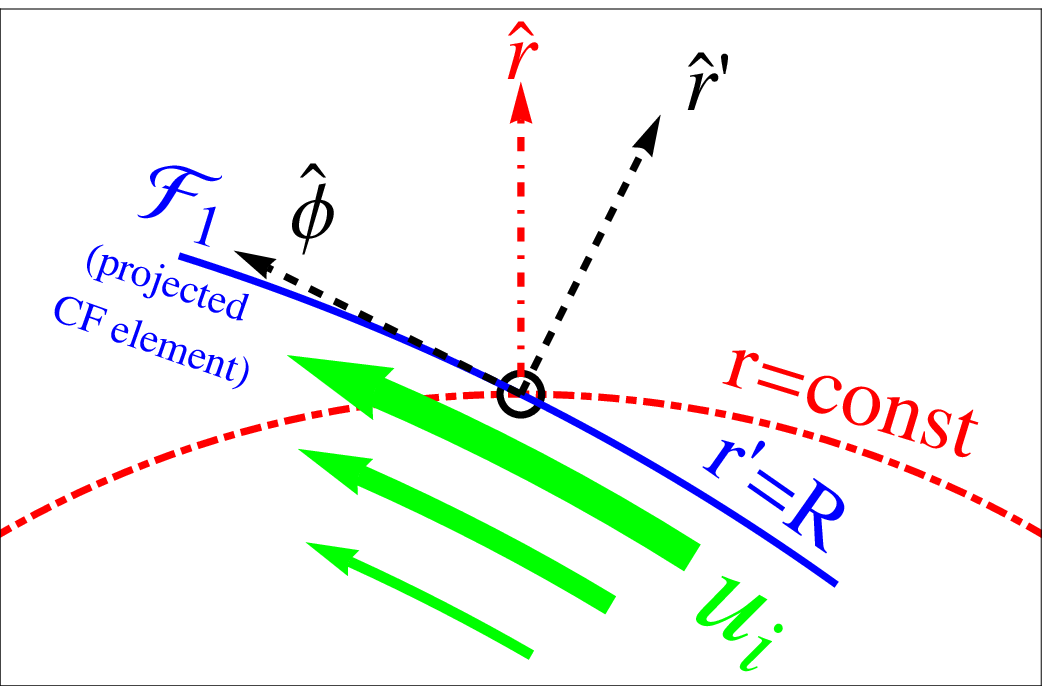}}}
\caption{
Local geometry of the CF and the inner flow.
The radius of curvature $R$ of the CF profile $\myF_1$ (solid curve) at a given position (circle) is typically slightly larger than the distance $r$ measured along $\unit{r}$ (dot-dashed) from the cluster center.
The flow (filled arrows, arbitrarily shown as an outflow) near the CF must parallel $\myF_2$ in a frame $S'$ (dashed axes) comoving with the CF.
The observed $M_g$ profile implies that shear flow extends far beneath the CF (filled arrows of different sizes).
\label{fig:CFIllustration}
}
\end{figure}

Consider an infinitesimal, two-dimensional piece $\myF_2$ of the CF plane, and an inertial ``CF frame'' $S'$ instantaneously comoving with $\myF_2$ perpendicular to the CF plane, as illustrated in Figure \ref{fig:CFIllustration}.
Let $\myF_1$ be an arc in $\myF_2$ parallel to the local flow beneath the CF. We may approximate $\myF_1$ as a segment of a large circle with radius $R$. Consider the spherical coordinate system $S'=(\myr,\myth,\myph)$ with $\myF_1$ lying at radius $\myr=R$ and colatitude $\myth=\pi/2$. Its origin, the center of the circle, is typically close but not necessarily coincident with the cluster center.

The $\myr$ component of the velocity $\vect{u}$ measured in $S'$ must vanish at $\myF_2$, $u_{\myr}=0$, as $\myhatbr$ is perpendicular to the discontinuity. The $\myr$ component of the acceleration is continuous across $\myF_2$, $\Delta (\pr_t u_{\myr})=0$.
Hence, taking the difference in the $\myr$ component of the Euler (momentum) equation between the two sides of $\myF_2$ yields
\begin{eqnarray} \label{eq:momentum_r_CF}
\Delta(u^2)/\myr = \Delta(\rho^{-1}\pr_{\myr} P)  \coma
\end{eqnarray}
where the pressure $P=\myPth + \myPNT$ may include thermal and nonthermal components, $u^2\equiv \vect{u}\cdot\vect{u}=u_\myph^2+u_\myth^2$, and we assumed slow changes in the CF pattern, $\vect{u}\cdot\grad{u_\myr}\ll u^2/\myr$.
Note that although $\vect{u}$ is confined to the CF plane, it does not have to be parallel on the two sides of $\myF_2$.

Equation (\ref{eq:momentum_r_CF}) indicates that the observed discontinuity $\Delta(\rho^{-1}\pr_\myr \myPth)>0$ must be balanced by a discontinuously larger centripetal acceleration $u^2/\myr$ beneath the CF.
The discontinuity cannot be plausibly attributed to $\myPNT$, because the continuous $\myPth$, noncontinuous $\grad{\myPth}$ combination observed would then require predominantly nonthermal pressure at large, sometimes $r>50\kpc$ radii, which is unlikely  \citep[\eg ][]{ChurazovEtAl08}.
Henceforth we neglect $\myPNT$.

We conclude that shear flow across CFs is the most natural interpretation of the observations. Thus
\begin{eqnarray} \label{eq:pressure_term_discontinuity}
u_\myin^2-u_\myout^2 = \Delta (u^2) \simeq (k_B/\mu)\Delta (\mypP' T) \simeq -r \Delta a_r > 0 \coma
\end{eqnarray}
where $\mypP'\equiv \pr\ln P/\pr\ln \myr\simeq (\myr/r)(\unit{r}\cdot\unit{r}')\mypP \simeq \mypP$.
Hence, CFs are tangential discontinuities, with shear and faster tangential flow beneath the CF, $u_{\myin}>u_{\myout}$.
In particular, $\Delta u\leq\mydu\leq u_\myin$, where $\mydu\equiv [\Delta(u^2)]^{1/2}$ is measurable.

In terms of the local sound velocity, Equation~(\ref{eq:pressure_term_discontinuity}) becomes
\begin{eqnarray} \label{eq:u2_jump}
\Gamma \Delta(u^2) = (q^{-1}\mypP'_{\myin}-\mypP'_{\myout})c_{\myout}^2 = (\mypP'_{\myin}-\mypP_{\myout}'q)c_{\myin}^2 \fin
\end{eqnarray}
Typically the $P$ slopes are $\mypP_\myin \gtrsim -0.7$, steepening to $\mypP_\myout\lesssim -1$ above the CF, so a subsonic flow in the CF frame ($u_\myin<c_\myin$) requires $q\lesssim 2.4$. Indeed, $q\lesssim 2.5$ is observed in CC CFs \citep{OwersEtAl09}.
Assuming the subsonic flow and continuous pressure, Equation~(\ref{eq:u2_jump}) also imposes an upper limit on the velocity above the CF, $\mach_{\myout}^2\equiv (u_o/c_o)^2<\myq^{-1}+\mypP'_{\myout}/\Gamma - \mypP'_{\myin}/(\Gamma\myq)$.

As an example, consider the CF in A1795. The $M_g$ discontinuity was previously interpreted as reflecting radial acceleration, $a\sim 2\times 10^{-8} \cm\se^{-2}$. Interpreted as tangential flow, this corresponds to $\mydu\simeq (ar)^{1/2}\sim 750 \km \se^{-1}$, uncertain to within a factor of $\sim 2$. This velocity difference constitutes a considerable fraction of the sound velocity, $\Delta u_2\simeq 0.67c_{\myin} \simeq 0.55c_{\myout}$. It places a lower limit $\mach_{\myin}>0.67$ on the (tangential) Mach number beneath the CF, and an upper limit $\Delta u<0.67c_{\myin}$ on the (nonzero) shear across it. Assuming $\mach_{\myin}<1$, it also imposes an upper limit $\mach_{\myout}<0.61$ on the flow above the CF.

Note that the tangential flow inferred from $\Delta a_r\neq 0$ may coexist with perpendicular (radial) motion of the CF through the ICM, inferred from an apparent pressure discontinuity, for example in RX J1720+26 \citep{MazzottaEtAl01}.

The $M_g(r)$ profile near a CF can be compared to other, equidistant regions in the core, preferably on the opposite side of the cluster, that do not harbor discontinuities.
In both cases examined, A1795 (Figure \ref{fig:CFJump}, inset) and RX J1720+26, $M_g(r>r_0)$ is similar on both sides of the cluster above the CF radius $r_0$, but inertia appears to be missing ($M_g$ is too small) in most of the volume beneath the CF.
In the tangential flow picture, the missing inertia region beneath the CF corresponds to an extended tangential bulk flow.
Consider again the CF lying $r_0 \simeq 85\kpc$ south of the center of A1795.
Here, inertia appears to be missing in a considerable, $0.7\lesssim r/r_0<1$ region, involving $\sim 1/2$ of the gas in that sector.
There is no evidence for inertia discrepancy below $\sim 0.6r_0$. Hence, plasma within the discrepant radii range cannot be uniformly flowing; shear must be present.

The existence and extent of the shear flow region are also gauged by nearly radial optical (and sometimes X-ray) filaments that lie below some CFs.
In A1795, a pair of filaments extend out to $\sim 0.6 r_0$ \citep[][and references therein]{McDonaldVeilleux09}, just beneath the shear region.
In M87, the South-West filaments \citep{FormanEtAl07} extend out to $\sim 0.6r_0$. They are nearly radial out to $\sim 0.4r_0$, but have some curvature at $0.4r_0\lesssim r \lesssim 0.6r_0$ which may be driven by shear.
The (sometimes curved) termination of filaments just below the shear region suggests that they may have been eroded by the shear.

\section{Magnetic shear amplification}
\label{sec:MagneticAmplification}

An independent argument implying shear across CFs stems from their remarkable sharpness and stability.
Stationary CFs are stable against Rayleigh-Taylor instabilities, as the inside plasma has lower entropy.
However, particle diffusion, heat conduction, and instabilities such as Kelvin-Helmholtz (KHI) and Richtmyer-Meshkov, could potentially broaden and deteriorate the CF on short, subdynamical time scales.
In contrast, observations indicate that CFs are not only stable over cosmological time scales, but in fact the transition in thermodynamical properties subtends a small fraction of the Coulomb mean free path of protons \citep{VikhlininEtAl01A, MarkevitchVikhlinin07}.

This suggests that transport across CFs is magnetically suppressed.
Such suppression is expected in CFs that move through the ICM, due to draping of magnetic fields and possibly the formation of a magnetic barrier (plasma depletion layer), as found in observations of planetary magnetospheres and in simulations \citep[][and references therein]{Lyutikov06}.
However, the processes that suppress transport in the more common, stationary type of CFs, were thus far unknown.
Here we argue that shear magnetic amplification near CFs naturally sustains the strong magnetic fields parallel to the CF needed to suppress radial transport.

Consider first the limit where CFs are infinitely sharp tangential discontinuities.
The magnetic field perpendicular to such a CF must vanish, and there is no shear magnetic amplification.
However, in this limit the stability against KHI inferred from observations requires a strong magnetic field $\vect{B}$, as shown by \citet{VikhlininEtAl01B} for the merger CF in A3667.
Namely, the shear velocity difference must be smaller than an effective Alfv\'{e}n velocity \citep[][ \S53]{LandauLifshitz60_EM},
\begin{eqnarray} \label{eq:KHI1}
\left( \frac{\Delta u}{2} \right)^2 < \frac{\bar{B}^2}{4\pi\bar{\rho}} \simeq \frac{3(1+\myq)}{5\bar{\beta}_i} c_i^2 \coma
\end{eqnarray}
and in addition
\begin{eqnarray} \label{eq:KHI2}
\left( \vect{B}_{\myin} \times \vect{B}_{\myout} \right)^2 \geq 2\pi \bar{\rho} \left[  \left(\vect{B}_{\myin} \times \Delta{\vect{u}} \right)^2 +\left(\vect{B}_{\myout} \times \Delta{\vect{u}} \right)^2 \right] \coma
\end{eqnarray}
where $\bar{B}^2\equiv (B_{\myin}^2+B_{\myout}^2)/2$, $\bar{\rho} \equiv 2\rho_{\myin}\rho_{\myout}/(\rho_{\myin}+\rho_{\myout})$, and $\bar{\beta}\equiv \myPth/P_{\bar{B}}=8\pi nk_B T/\bar{B}^2$, with $n$ being the particle number density.

Equation (\ref{eq:KHI1}) imposes a lower limit on $\bar{B}$,
\begin{eqnarray}  \label{eq:KHI3}
\bar{B} > B_{min} & \equiv & \sqrt{\frac{2\pi\rho_i}{1+\myq}} |\Delta u| = \sqrt{\frac{5}{12(1+q)}} \frac{|\Delta u|}{c_i} B_{eq,\myin} \nonumber \\
& & \simeq 17 \mach_i \sqrt{ \frac{3n_i k_B T_i/(1+q)}{50\eV\cm^{-3}} } \muG \fin
\end{eqnarray}
The large shear velocities derived in \S\ref{sec:DifferentialRotationAcrossCF} thus imply large $\bar{B}$, up to $\sim40\%$ of its equipartition value $B_{eq}$ for which $\beta=1$.
An asymmetry $B_{\myin}/B_{\myout}\neq 1$ would require $\vect{B}$ well aligned with $\Delta \vect{u}$.
In the case $B_{\myin}=B_{\myout}=B$, $\alpha_\myin\neq\alpha_\myout$,
\begin{eqnarray}  \label{eq:KHI4}
B > \sqrt{\frac{2(\sin^2\alpha_{\myin}+\sin^2\alpha_{\myout})}{\sin^2(\alpha_{\myin}-\alpha_{\myout})}} B_{min} \coma
\end{eqnarray}
with $\alpha$ being the angle between $\vect{B}$ and $\Delta \vect{u}$, so stronger fields are needed if misaligned with the shear.

A magnetic, plasma depleted barrier may form.
As long as its temperature does not greatly exceed $T_\myout$, it remains strongly magnetized.
Note that the corresponding density drop could possibly be observed more easily here than in merger-type CFs, as there is no confusion with a stagnation region.
As shown below, equipartition fields are expected only very close to the CF, so $\myPNT$ can still be neglected at $\gtrsim\kpc$ distances, as assumed in \S\ref{sec:DifferentialRotationAcrossCF}.

How did strong magnetic fields form along stationary CFs, and how are they sustained over cosmological time scales?
Perpendicular transport suppression could preserve such fields practically indefinitely in the absence of
large perturbations (\eg mergers) and disruptive kinetic plasma effects.
Hence, they could be remnants of an early stage where the CF was formed.
But if such initial fields are absent or decay, shear driven magnetic amplification can naturally generate or replenish $B_\phi$, by stretching $B_\myr\neq0$ magnetic structures advected by the flow via line freezing, $d(\vect{B}/\rho)/dt\simeq[(\vect{B}/\rho)\cdot\nabla]\vect{u}$.

Consider a weakly magnetized, $B\ll B_{eq}$ stage where the kinematic viscosity $\nu$ is a fraction $f_\nu$ of its Spitzer value.
The velocity discontinuity across such a CF is gradually smoothed out, forming a transition layer of thickness $\myl\sim (\nu t)^{1/2}\simeq 7 f_\nu^{1/2} T_4^{5/4}n_{-2}^{-1/2}(t/\mbox{Gyr})^{1/2}\kpc$ beneath the CF at time $t$, where $T_4\equiv (T_{\myin}/4\keV)$ and $n_{-2}\equiv (n_{\myin}/10^{-2}\cm^{-3})$.
This enables rapid shear magnetic amplification, on an e-fold time scale $\tau\sim \myl/u$. Equating the viscous and amplification times suggests that a magnetization layer of thickness
\begin{eqnarray}
\myl \sim N\nu/u \simeq 50 N f_\nu \mach_i^{-1} T_4^2 n_{-2}^{-1} \pc
\end{eqnarray}
develops over a very short time scale
\begin{eqnarray}
t \sim N^2\nu/u^2 \simeq 4\times 10^4 N^2 f_\nu \mach_i^{-2}T_4^{3/2}n_{-2}^{-1}\yr \coma
\end{eqnarray}
with $N$ being the growth factor. Typically $Nf_\nu<1$; $N\sim 1$ may suffice in the case of a weakening magnetic barrier.

The high $B$ estimates above hold only in the vicinity of CFs, where fast shear is measured.
As mentioned in \S\ref{sec:DifferentialRotationAcrossCF}, evidence suggests that the associated shear involves a fair fraction of the plasma beneath the CF.
Shear amplification may thus effectively magnetize the entire core, at least beneath the outermost CF.
For a simple model where the shear is constant in a layer of thickness $\Delta$ beneath the CF, the energy of the magnetic field is amplified in proportion to $(B_\phi/B_\myr)^2$, where
\begin{eqnarray} \label{eq:ShearAmplification}
\frac{B_{\phi}}{B_\myr} \sim t \pr_\myr u \sim 10 \mach_i T_4^{1/2} \left(\frac{\Delta}{10\kpc}\right)^{-1} \left(\frac{t}{10^8\yr}\right) \fin \,\,\,
\end{eqnarray}
Saturation of this process depends on the flow details. For example, in a spiral flow there is a characteristic cutoff time when magnetic structures leave the shear region.

We have seen that CFs are strongly magnetized, and that the associated shear flow can magnetize much of the plasma beneath them.
Magnetic processes should therefore be enhanced below, and in particular near CFs.
This naturally explains the recently discovered coincidence between CFs and the edges of radio minihalos (MH) \citep{MazzottaGiacintucci08}; see \citet{KeshetLoeb10}.

\section{Discussion}
\label{sec:Discussion}

We interpret deviations from hydrostatic equilibrium previously observed near CFs as evidence for centripetal acceleration.
This reveals a tangential shear flow, nearly sonic at the CFs and extending well beneath them.
Such behavior is shown to be common among CC CFs, as illustrated in Figure \ref{fig:CFJump}.
The measured discontinuity $\Delta (u^2)$, given by Eqs.~(\ref{eq:pressure_term_discontinuity}-\ref{eq:u2_jump}), constrains the shear and inside flow, $\Delta u\leq \mydu \leq u_i$, and imposes an upper limit on $\mach_\myout$. The apparent CF stability against KHI suggests near-equipartition magnetic fields along the CF (Eqs.~\ref{eq:KHI3}-\ref{eq:KHI4}); this could manifest in a shear-amplified, $\lesssim 50\pc$ magnetic layer.
Shear magnetic amplification could rapidly magnetize the entire core up to the CFs, and explain some of the observed properties of radio MHs.

In our model, core CFs are part of extended tangential discontinuities seen in projection.
They directly trace both the orientation and approximate magnitude of bulk flows in the ICM.
CFs are usually nearly concentric, so $\vect{B}$ should be nearly tangential and the polarization radial.
Note that this magnetic configuration differs from the predictions of magnetic compression models, where $\vect{B}$ is typically isotropic \citep[\eg][]{GittiEtAl02} or radial \citep{SokerSarazin90},
but is similar to the saturation of heat flux driven buoyancy instabilities in the core \citep{Quataert08}.
In MHs, we predict detailed morphological correlations with X-ray gradients, which trace projected CFs, and radio polarization parallel to these gradients.

Such fast, extended flows would modify the energy budget of the core and could alter its structure, for example by distributing heat and relaxing the cooling problem.
\cite{ChurazovEtAl03} pointed out that the absence of resonant features in the X-ray spectrum in Perseus indicates motions with at least $\mach=1/2$ in the core. Although initially interpreted as turbulent motion, this may reflect the bulk motions discussed here.
The presence of such flows could be tested, for example using high resolution spectroscopy and future X-ray polarization measurements \citep[\eg][]{SazonovEtAl02}.

\acknowledgements
We thank W. Forman and O. Cohen for useful discussions.
UK acknowledges support from NASA through Einstein Postdoctoral Fellowship grant number PF8-90059,
awarded by the Chandra X-ray Center, which is operated by the Smithsonian Astrophysical Observatory for NASA under contract NAS8-03060.
This work was supported in part by NASA contract NAS8-39073 (MM), by an ITC fellowship from the Harvard College Observatory (YB), and by NSF grant AST-0907890 (AL).

\end{document}